\documentclass[aps,prl,10pt,reprint,groupedaddress,superscriptaddress]{revtex4-2}
\usepackage{graphicx}% Include figure files
\usepackage{dcolumn}% Align table columns on decimal point
\usepackage{bm}% bold math
\usepackage[nointegrals]{wasysym}
\usepackage[unicode=true,colorlinks=true]{hyperref}
\hypersetup{linkcolor=blue,citecolor=blue,urlcolor=blue}
\usepackage{graphicx}
\usepackage{epstopdf}
\usepackage{color}
\usepackage{amsmath}
\usepackage{amsfonts}
\usepackage{amssymb}
\usepackage{bbm}
\usepackage{wasysym}
\usepackage{csquotes}
\usepackage[normalem]{ulem}
\DeclareMathOperator{\Tr}{Tr}
\usepackage{booktabs}
\usepackage{braket}

\usepackage{dsfont}
\usepackage{slashed} 

\usepackage{tikz}

\definecolor{orange}{RGB}{255,127,0}
\definecolor{blue2}{RGB}{33,114,173}

\renewcommand{\Re}{\operatorname{Re}}
\renewcommand{\Im}{\operatorname{Im}}

\newcommand{\iu}{\mathrm{i}} % Math symbol for \sqrt(-1)
\newcommand{\bvec}[1]{{\bm{#1}}}

\newcommand{\Ztwo}{\mathbb{Z}_2}

\newcommand{\ph}{\mathrm{\ph}}

\begin{document}
% \preprint{APS/123-QED}

% Force line breaks with \\Haldane impurity Chern number reversal and a zero-disorder induced tri-critical point

\title{
Dichroic Raman probes for chiral edge modes 
}

\newcommand{\TUM}{\affiliation{Technical University of Munich, TUM School of Natural Sciences, Physics Department, 85748 Garching, Germany}}
\newcommand{\MCQST}{\affiliation{Munich Center for Quantum Science and Technology (MCQST), Schellingstr. 4, 80799 M{\"u}nchen, Germany}}
\newcommand{\Imperial}{\affiliation{Blackett Laboratory, Imperial College London, London SW7 2AZ, United Kingdom}}
\newcommand{\Cologne}{\affiliation{Institute for Theoretical Physics, University of Cologne, 50937 Cologne, Germany}}

\author{Avedis Neehus} \TUM \MCQST
\author{Johannes Knolle} \TUM \MCQST

\date{\today}% It is always \today, today,
             %  but any date may be explicitly specified

\begin{abstract}
The identification and manipulation of charge-neutral fractionalized quasi-particles, in particular chiral edge modes (CEM), is a long-standing quest in physics. Remarkably,  the microscopically mediated  interaction between light and charge-neutral excitations in Mott-Hubbard insulators can take an identical form to the Raman coupling between light and particles with electric charge. However, since CEMs are Raman-inactive due to conservation of lattice momentum, Raman probes have been deemed unsuitable for their identification. Here, using the Kitaev quantum spin liquid (KSL) as an illustrative example, we demonstrate that the long-range correlated disorder inherent to a closed edge can lead to a Raman circular dichroism (RCD) signal that avoids suppression by linear and angular momentum selection rules, and exhibits a dependence on experimentally tunable length and energy scales that are characteristic of CEM. Having calculated the low-frequency RCD response of generic KSL, we argue that the interaction of the chiral matter fermion with the $\Ztwo$ boundary charge leaves a unique fingerprint of the KSL via the anisotropic Zeeman field dependence.
\end{abstract}
\maketitle
{\em Introduction.---} The notion of bulk-boundary correspondence, whereby a topological index  associated to the bulk of a system is related to the existence of zero modes at the boundary, has proven invaluable across a wide breadth of modern physics \cite{Teo10}. A salient example of this phenomenon is the (anomalous) quantum Hall effect \cite{Klitzing80}, where the Chern number of a bulk insulator predicts the number of CEM at the Fermi level that manifest in a quantized charge Hall conductivity \cite{TKNN82}. If, however, the Chern number is associated with charge-neutral particles that couple only to an emergent $U(1)$ field, e.g., in p-wave superconductors or chiral spin liquids, the experimental signatures of edge modes are more subtle. Although a \textit{thermal} Hall conductivity of the neutral particle systems could be quantized, the contributions from and coupling to other low energy degrees of freedom (d.o.f.), i.e. phonons, easily spoil clear quantization~\cite{vinkler2018approximately,ye2018quantization}. Indeed, unambiguous verification of topologically protected edge modes, in particular charge-neutral ones like Majorana zero modes (MZM), has been a long-standing quest of the condensed matter community~\cite{wu2024charge}, not in the least because of its potential applications in the field of quantum computing \cite{Yazdani23}.\\
In this work, we discuss the theory of Raman scattering as a probe of CEM, with a particular focus on the Kitaev spin-liquid (KSL).
Earlier works on this topic had concluded that the main obstacle to Raman spectroscopy, a zero-momentum-transfer process, of CEM lies in the fact that they require a finite momentum transfer $q$ to be excited by a local operator \cite{Johannes16, Nagaosa21}. He and Nagaosa \cite{Nagaosa21} proposed enhancing the signal with focused light, which would enable finite-momentum-transfer processes, and found a purportedly universal exponent for the Raman shift dependence of the Raman intensity. However, the thus generated signal is still of very low magnitude, especially considering that it would compete with the acoustic phonon contribution that also increases with $q$. Hence, it appears that the Raman probe can not solve the problem that plagues thermal Hall measurements, i.e., being masked by phonon signals. 
\begin{figure}
    \centering
    \includegraphics[]{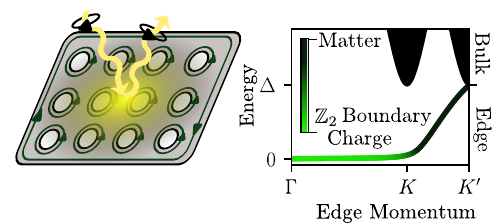}
    \caption{Sketch of the proposed protocol whereby the Raman intensity is measured as the shape and number of holes is varied, thereby tuning the surface to bulk ratio. The Raman intensity at Raman shift frequencies below the bulk gap $\Delta$ probes the chiral edge mode. As sketched on the right panel, in the KSL, the edge mode is a superposition of a chiral matter fermion with support at finite momenta between the Dirac points, and the $\mathbb{Z}_2$ boundary charge d.o.f.   }
    \label{fig:setup}
\end{figure}
Strikingly, we find that when considering $q=0$ Raman scattering on samples with curved edges, which is the realistic scenario, the concomitant introduction of a large translation symmetry breaking length scale $L$, and corresponding energy scale $E_L$,  allows for the simultaneous evasion of linear and angular momentum selection rules for transitions mixing the small $E< E_L$ and large $E>E_L$ energy scales. With an eye toward the experimental challenges of isolating the magnetic contribution to Raman intensity, we focus on Raman circular dichroism (RCD), which further suppresses phonon contributions and other background signals.

Furthermore, we point out that previous analysis of Raman scattering in the Kitaev spin liquid \cite{Nagaosa21,Johannes16, feldmeier2020local} neglected to consider the boundary charges related to the bulk $\Ztwo$ gauge d.o.f. that interact with the chiral edge mode deriving from the finite Chern number of the matter fermions~\cite{feldmeier2020local,zhang2025probing}. Accounting for this interaction shows that the low-energy Raman signal is, in fact, highly dependent on material-specific parameters. However, we show that the unique way in which the boundary charges couple to the Zeeman field provides a strong fingerprint of the underlying spin-liquid phase.
Meanwhile, evidence towards the existence of CEM comes not from a finite in-gap RCD signal itself, which we argue, in contrast to previous claims~\cite{Bostrom23,Lu22}, is not directly related to the modes chirality,   but rather its dependence on the energy scale $E_L$, which can be tuned by, e.g., the edges geometry and the Zeeman field. 
Given that the scale $L$ of a samples boundary is difficult to control and may lie in an unfavorable regime we propose a simple remedy; namely, to introduce additional edges by systematically nano-structuring holes into the sample, as established for antidot lattices with hole sizes below  ~100nm~\cite{neusser2010anisotropic}. Beyond enabling control over the edges shape, this procedure also helps attribute the low-frequency Raman signal to edge modes, as possible in-gap bulk contributions would be reduced by systematically increasing the surface-to-bulk ratio.

{\em Raman scattering.---}
We consider non-resonant Raman scattering in a Mott-insulator described by a Hamiltonian of spin operators $S$, $
    H^S = \sum_{ij} J^{\alpha \beta}_{ij}  S^\alpha_{i}S^\beta_{i}$
where, the leading Raman operator $R$ takes the Loudon-Fleury form\cite{Fleury68,Shastry90}  \begin{align}
    R^{p p^\prime } = \sum_{ij} \lambda (\bvec{e}^p \cdot \bvec{d}_{ij})(\bvec{e}^{p^\prime} \cdot \bvec{d}_{ij}) J^{\alpha \beta}_{ij} 
S^\alpha_{i}S^\beta_{j} \label{eq:Ramanop}
\end{align}
with $\bvec{e}^p$ the photon  polarization vector, $\bvec{d}_{ij}$ the distance between two sites and $\lambda$ a microscopic coupling constant.
Hence, if the spin Hamiltonian has an effective description in terms of local kinetic energy of particle operators, the Raman coupling is formally the same as for fermions that carry an electric charge, where it also takes the form $R_{ij} \sim \bvec{d}_{ij} \bvec{d}_{ij} H^e_{ij}$, which has been recently exploited to make a connection to an effective light-matter coupling~\cite{koller2025raman}. In this sense, our results for Mott-Hubbard insulators and phases with itinerant electrons are analogous.  
For concreteness, we focus on the  Kitaev honeycomb model \cite{Kitaev_2006,Khalul09}
\begin{align}
    H^K = \sum_{\langle ij \rangle_\gamma} J^{\gamma}_{ij} \sigma^\gamma_i \sigma^\gamma_j,
\end{align}
where $\sigma$ are Pauli operators for the spin 1/2 d.o.f. and $\gamma$ are assigned according to the direction of the nearest neighbor vector. The Majorana decomposition $
    \sigma^\alpha = ib^\alpha c, \quad c\prod_\alpha \iu b^\alpha = \iu$
factorizes the eigenspace of $H^K$ into flux sectors corresponding to the sign of the expectation values of the bond fermions $\langle ib_ib_j\rangle := u_{ij}$ along closed loops, and the matter d.o.f. fermions $c$. Choosing isotropic couplings $J^{\gamma}_{ij}=J$, fixing the ground state flux sector, and perturbatively taking into account an out-of-plane magnetic field $h$, the effective description is determined by the Hamiltonian 
\begin{align}
        H^f&=J \sum_{\langle i, j\rangle}\iu c_ic_j+  \kappa\sum_{\langle\langle i, j\rangle\rangle }  \iu \nu_{ij}  c_i c_j. \label{eq:haldane}
\end{align}
where $\nu =\pm$ is chosen positive for an orientation that goes clockwise around the center of a plaquette. The bulk spectrum of the above Hamiltonian has a topological spectral gap $ \Delta = 3\sqrt{3}\kappa \sim  h_xh_yh_z$. Therefore, we find an effective description in terms of Chernful fermion bands, which imply the existence of gapless chiral excitations at the edge.\\

\begin{figure}[!h]
    \includegraphics[]{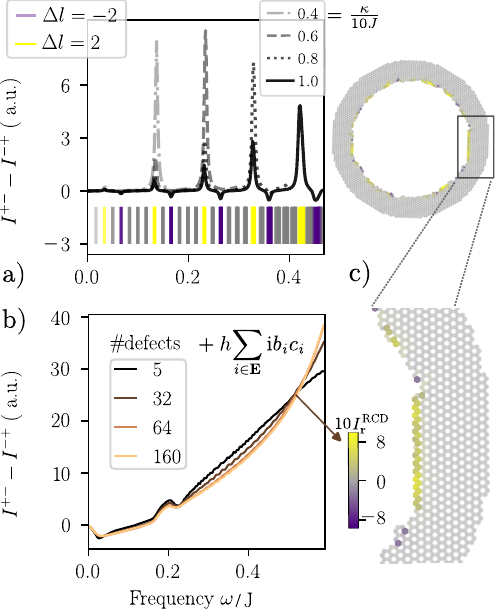}
    \caption{  Raman circular dichrosim   $I^{+-}-I^{+-}$ of a punctured disk of a perfect Kitaev material. Panel (a) shows the case with no bond-matter hybridization for various $\kappa$ up to the $\kappa$ dependent $\Delta$ for a hexagonal disk with $r_1=6$, $r_2 =56a$. The energetically allowed transitions are indicated by the vertical bars at the bottom, whose colors encode the angular momentum transfer as determined in the end matter. Panel (b) shows the case with bond matter hybridization $h=0.1J$ and $\kappa = 0.12J$ for various defect densities described by equation \eqref{eq:defect} on a circular $M=12$ disk with $r_1 =36a$ and $r_2 = 46a$. The RCD in the shown frequency regime is attributed to the inner edge, as illustrated in panel (c), which shows a spatially resolved contributions at $\omega = 0.5J$.}
    \label{fig:fig1}
\end{figure}
In principle, these boundary modes should manifest as a subgap low-frequency $\omega$ signal in the Raman intensity
\begin{align}
   I^{p p^\prime}(\omega)  = \int_{\mathbb{R}}dt  e^{i\omega t} \langle R^{p p^\prime }(t) R^{p^\prime p }\rangle_\beta ,
\end{align}
where $\langle \rangle_\beta$ is the thermal expectation value at inverse temperature $\beta$. 
However, for edges with translational invariance, the Raman edge signal is impeded because it only probes $q =0$ excitations and, for $\omega $ below the bulk gap $\Delta$, CEM cannot be exited without transfer of linear momentum. While this problem can be circumvented by introducing disorder, it is usually short-range correlated, so that low-energy modes still have approximate translation invariance. Furthermore, the disorder strength should be small compared to $J$, as it would otherwise also spoil the bulk spectral gap. 
Alas, real samples are finite and have a single \emph{curved} edge; hence, translation invariance is intrinsically broken by the geometry of the sample on large length scales. Assuming for now that the geometry of the sample preserves the point group symmetry of the lattice, one may instead consider angular momentum $l$. Since, relative to the sample's normal vector, the RCD absorbs a photon with a chirality
opposite to the one it emits, the angular momentum selection rule is $\Delta l = \pm 2$.

In Fig. \ref{fig:fig1} we show the Raman circular dichroic response $I^{RCD}=I^{+-}-I^{-+}$ with $\bvec{e}^\pm = (1,\pm i)$ of a disorder free punctured disk with inner radius $r_1$ and outer radius $r_2$, demonstrating that the operative selection rule is conservation of (lattice) angular momentum $l$, which is extracted from the (projective) action of origin centered lattice rotations $U_{C_6}C_6$,  where $U$ is a gauge transformation, on the single particle eigenstates $\ket{n,l}$ of the Hamiltonian, i.e., $U_{C_6}C_6 \ket{n,l} =e^{i 2\pi\frac{l+1/2}{6}} \ket{n,l}$, with the additional $1/2$ coming from $(U_{C_6}C_6)^6 = -1$ in the vison free gauge-sector. Accounting for the possibility of a vison pair forming on the two edges, we also determined the projective representation of $C_6$ in the presence of an edge connecting gauge string. The spectrum of the projective rotation is now shifted such that $(U_{C_6}C_6)^6 = 1$, which leads to a shift of the Raman spectrum proportional to  $r^{-1}$. We find that the zero-mode subspace has a component with $l=3$ and one with $l=0$ such that a Raman transition involving a MZM pair requires an angular momentum transfer of $|\Delta l| = 3$, and is thus unobservable. 

Having established the operative selection rule in this disk geometry, we now discuss the interplay between approximate momentum and angular momentum conservation that allows for a finite edge mode Raman signal. 
Indeed, consider that the low energy dispersion relation for a chiral edge mode is linear with velocity $v$, $E(l) \propto  \hbar v\frac{l+1/2}{l_{max}}$ \cite{Stone04}. In the continuum limit the system possesses a $U(1)$ rotation symmetry and, therefore, by angular momentum conservation, one would only expect a single peak at $E(\pm 2) \propto l_{max}^{-1}\propto r^{-1}$ where the sign depends on the sign of $v$. A crucial point is that because lattice angular momentum is only defined modulo some integer $n$, the selection rule becomes $\Delta l \mod n = \pm 2 $ and one obtains a signal in the whole frequency range. To determine the energy scale that interpolates between the discrete and continuous rotation symmetries, consider the geometry of the sample to be described by a polygon with $M$ sides inscribed on a circle of radius $r$, which defines a length scale $L = 2\pi r/M$. Using the length scale $L$, we may define the energy scale $E_L = \hbar v/L$. For modes with $ E \ll E_L$, the continuum description is appropriate; on the other hand, modes with $E \gg E_L$ only respect lattice rotation symmetry. Similarly, an inverted relationship holds for linear momentum selection rules \footnote{  Since $R$ is a sum of operators $R_x$ each with local support, we may consider the matrix element of $R_x$, which depends on $ \psi_n(x) = \braket{\bm x|n}$. Modeling two linear segments connected by a corner as two decoupled linear chains $o$ and $d$ of length $L$ with periodic boundary conditions, connected by a local potential $V$  placed at the origin, the correction to the translationally invariant eigenstate $\psi^o_n$ of the $o$ chain is given by 
 \begin{align}
     \delta\psi^o_n( x) =  G(E_n,x,0) V(0) \psi^d_n(0). 
 \end{align}
where $G$ is the one-particle Greens function. For the chiral edge mode  $G(E,x, 0) \propto e^{iEx/{\hbar v}}$ and  assuming  $|\psi^d(x)|^2$ is uniform, $\psi(0) \propto L^{-1/2}$, $R_x \propto \delta\psi^o_n( x) \propto e^{iEx/{\hbar v}}/\sqrt{L} $ .  Then the Raman intensity  due to one strip goes as $ I \propto \frac{L}{L}\int^L_0  dx e^{2iEx/{\hbar v}} $. Thus, for long wavelength modes, i.e., $EL \ll \hbar v$ the correction from the translation symmetry breaking $V$ leads to a finite Raman signal. On the other hand, when $EL \gg \hbar v$ ,  the correction starts oscillating and  averages out. },  where it is avoided by modes with large wavelength, i.e., $E\ll E_L$ while the Raman signal of small wavelength modes  with $E \gg E_L$ is suppressed. Now, because Raman scattering is a two body process, above the frequency $E_L$ there are matrix elements that mix the $E\ll E_L$ and $E \gg E_L$ scales and do not obey angular nor linear momentum selection rules, see also the End Matter for further numerical analysis of the Raman matrix elements.

We can see this mechanism at play in Fig.~\ref{fig:fig1}. Since the Raman matrix elements involving energies $E_n$ and $E_m$ grow with their respective distance from $E_L$, the magnitude of the RCD grows with $\omega$. By lowering $\kappa$, the bulk gap decreases, but also $v$ and $E_L$ decrease, such that the RCD for a fixed frequency can increase even though the Raman operator is proportional to $v$. This breaks with the usual expectations that the RCD should decrease with the time-reversal breaking parameter, i.e.,  $\kappa$, being lowered, and is a consequence of the lifting of selection rules which forbid a Raman signal specifically for CEM \footnote{We do not claim that $I_{RCD}$ is monotonically decreasing with $\kappa$ for all $\kappa$. However, the fact that a decreasing regime exists appears to be generic, as it is also true in the general setting with disorder and $h\neq0$.}.

Given that the RCD is sensitive to a state's chirality in the sense that it depends on a state's angular momentum, one might conjecture that it has a relation to the Chern number \cite{Bostrom23} or the chirality of an edge mode \cite{Lu22}. However, we find that neither the bulk nor the edge RCD can be used to infer the topological index, as further explained in the Appendix.

\begin{figure}[!h]
    \includegraphics[]{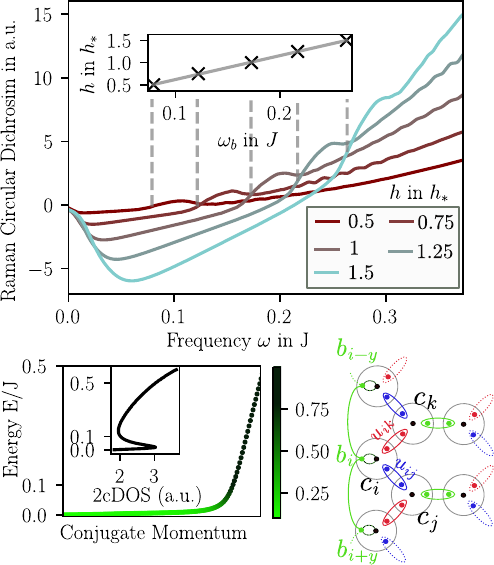} 
     \caption{$I^{RCD}$ with $\kappa = 0.12J$, $r_1 = 36a$ and  $r_2 = 46a$ for different values of the hybridization term written in multiples of $h^* = 0.1J$. The inset shows the linear relationship between $h$ and the threshold frequency $\omega_b$, which is determined from the local maximum of $\partial^2_\omega I(\omega)$ in the low frequency range. The bottom-left panel shows the low-energy dispersion, assuming it is monotonous. The color coding indicates the corresponding degree of matter character $\psi^n_c$, ranging from green (low) to black (high). The inset shows the 2cDOS, eq. \ref{eq:matter2DOS}. The $\Ztwo$ boundary charges arise as unpaired bond fermions as sketched in the bottom right panel, where we also indicate the possibility of a boundary hopping term that may be induced by non-Kitaev interactions. }
    \label{fig:fig3}
\end{figure}

{\em  Boundary hybridization.---}In addition to the chiral edge mode, the Kitaev spin liquid has another edge mode related to its nature as a deconfined gauge theory, namely sites which do not connect to three neighbors have an unpaired bond-fermion $b$ contributing an additional zero energy mode to every such site at the edge, see Fig. \ref{fig:fig3}. 
We include the bond-matter fermion hybridization up to first order in $h$:
\begin{align}
    H^h = H^f+ \sum_{i\in \mathbf{E}} h^{\gamma_i}\iu b_ic_i,
\end{align}
where $\mathbf{E}$ denotes the set of sites that have only 2 nearest neighbours and $\gamma_i$ is the label for the unpaired bond at site $i$. Note that here we understand $h$ as an appropriately rescaled Zeeman field.  
Importantly, the bond-matter fermion hybridization does not directly affect the Raman operator as the dipole factor $d_{ij}$ is zero for the Zeeman term. However, it does change the structure of the eigenstates as shown in Fig.~\ref{fig:fig3} where we indicate the degree of $c$ character by computing $\psi^n_c =  \Tr P_c \ket{n} \bra{n}$, with $P_c$ projecting to the matter sector. We observe that, up to the hybridization scale $h$, the low-energy states are predominantly of $b$-fermion character. To estimate the impact on the Raman signal, we also calculated the matter projected two-particle density of states.
\begin{align}
    \text{2cDOS} = \sum_{nm} \psi^n_c \psi^m_c \delta(\omega-(E_n+E_m)). \label{eq:matter2DOS}
\end{align}
While the 2DOS is relatively large in the low frequency range due to the small bandwidth of the bond fermions, the 2cDOS is suppressed by the predominant $b$-fermion character in this regime.
In addition to the bond fermions, we also consider the effect of two types of disorder. First, we draw $J_{ij}$ from a Gaussian distribution with mean $J$ and standard deviation $0.1J$. 
Secondly, we impose a roughening of the edge, which is achieved by selecting $N$ sites according to the probability distribution 
\begin{align}
    P(i) = \frac{1}{Zd_i}\exp{-\frac{(d_i-r_1)}{\xi}} \label{eq:defect}
\end{align}
where $d_i$ is the radial coordinate of the site $i$, $r_i$ the inner radius and $Z$ a normalization factor.  Then, we draw a circle around each selected site with radii $d_i-r_1$ and delete all sites that are inside the circle and have an origin centered radial coordinate smaller or equal to the corresponding $d_i$. We fix $\xi = 2a$ and $h^\gamma = h$. Lastly, we project out the edge modes localized at the outer edge. 
The resulting Raman signal is shown in panel b) of Fig.~\ref{fig:fig1}. Notably, the Raman spectrum is highly robust to shape disorder, reflecting the topological nature of the edge modes that protect it from localization. Furthermore, the Raman signal above $2h$ is enhanced. Referring to our previous discussion on the role of the energy scale $E_L\propto v$, we note that the relationship $v\propto \kappa$ is not universal. In particular, the coupling of the matter to the flat $b$ band may lower the effective velocity at low energies and hence the scale $E_L$.

Fig. \ref{fig:fig3}  shows the Raman circular dichroic response for the chiral Majorana modes of the KSL averaged over $500$ disorder realizations for various $\alpha$. As expected, an increase of $h$ (with constant $\kappa$) gives rise to a larger Raman intensity through an increased 2cDOS. The low-frequency peak in the 2cDOS is reflected in a peak with a sign opposite to the Raman edge signal of excitations above $2h$. Besides a shift in intensity,  the Zeeman field also 
affects a frequency shift of the peculiar features in the Raman signal that scales linearly in $h$. 

{\em  Non-Kitaev interactions---} Departing from the fine-tuned Kitaev limit, we now take into account weak time-reversal symmetric flux-off diagonal perturbations $\Gamma_i$ which endow the bond fermions  with a dispersion to order $ t \sim O(h\Gamma_i) $ \cite{zhang2025probing}:
\begin{align}
    H^{ht} = H^f+ \sum_{i\in \mathbf{E}} h^{\gamma_i}\iu b_ic_i+\sum_{ij\in \mathbf{E}} t_{ij}\iu b_ib_j \label{eq:dispdep}.
\end{align}
In principle, the non-Kitaev terms $\Gamma_i$ can also couple to $b$ without $h$, however for zig-zag type edges theses terms can be ignored, as also demonstrated numerically in the Appendix.  This is because, time-reversal symmetric perturbations couple opposite sublattices and the edge occupations have a strong sublattice imbalance depending on which site has a missing bond. Taking into account that the edge is made up of segments where the $b$ fermions reside either on the $A$ or the $B$ sublattice, the additional perturbations can only operate across these sublattice-domain walls, making the Zeeman field the dominant hybridization factor. The sublattice polarization is also the reason that the Raman operator \eqref{eq:Ramanop} does not couple to $b$, and the non-Kitaev vertices can be ignored below the flux gap.   This situation would be different in the regime of resonant Raman scattering \cite{Brent16, Johannes16}, in which case there is a Raman vertex that couples equal sublattices and thus also to the bond fermions. While the higher-order Raman vertex does not contribute to the RCD, the boundary charges lead to a greatly enhanced $I^{++}$, particularly as $h$ is lowered and the low-energy DOS diverges, as demonstrated in the Appendix.  

Even for the Raman operator, Eq.~\eqref{eq:Ramanop}, the $b$ hopping term still has a strong influence on the low-energy matrix elements of the RCD, as shown in the Appendix, where we observe that the low-energy behavior is highly dependent on the form of $t$. The only common denominator among these Raman spectra is their strong dependence on the Zeeman field, which affects both the hybridization and the hopping term. Thus,  given a method to engineer zig-zag type edges with, e.g., more missing $\gamma =x,y$ type bonds than $\gamma =z$ ,  the bond directional interaction of the boundary charges is reflected in a large anisotropy of the Raman signal with respect to the Zeeman field \footnote{For a mix of zig-zag and armchair type edges where $h$ independent hybridization/dispersion terms of the boundary charges become significant, the  anisotropy with respect to $h$ is still present but decreased.}. Therefore, while the low-frequency Raman signal itself is non-universal, Raman scattering still provides a signature of the underlying quantum spin liquid.

The non-Kitaev interactions also provide a pathway to change the value of $h$ independently of the matter gap, as the three-spin term responsible for the bulk gap is generated to order $h \Gamma_i$, hence losing the constraint of maintaining the product $h_xh_yh_z$. Since both the bond-fermion hopping term and the mass term appear to order $h \Gamma_i$ they could have similar magnitudes in principle, however, we find that the mass gap, which gets an additional factor $3 \sqrt{3}$ relative to $\kappa$, is generically larger then the bond-fermion bandwidth; the leading perturbations, the symmetric $\Gamma$ and anti-symmetric $\Gamma^\prime$ off-diagonal exchange, are respectively responsible for $t\propto 4\Gamma/\Delta_v$, $\kappa \propto 4 \Gamma^\prime/\Delta_v$, where $\Delta_v$ is the flux gap, but it is expected that $\Gamma/\Gamma^\prime <2$ \cite{moeller25}. 

\emph{Conclusion---} We have shown that while the Raman signal for CEM is suppressed by both linear and angular momentum selection rules, on a rounded sample edge  Raman transitions appear that obey neither selection rule if the energy scale $E_L$, introduced by the geometry of the sample in combination with the edge modes velocity, is finite and lies below the bulk gap. Therefore, in contrast to earlier predictions, CEM can contribute to an in-gap Raman {circular dichroism} signal. Although the RCD does not directly probe the edge modes' chiral nature, the dependence of the Raman signal on the scale $E_L$ can be used to differentiate it from other sources that may contribute to the in-gap RCD. For example, the RCD can increase as the Zeeman field is lowered. While the geometry of the sample as a whole may be difficult to control, the geometry dependence of the edge signal may be probed using nano-structured holes, which also give an additional evidence for the edge-mode character of the low energy signal, as potential bulk contributions  are reduced by increasing the surface to bulk ratio.
As a proof of principle, this technique could first be applied to conventional Chern insulators using well-established methods. Similarly, chiral magnon edge modes should have similar signatures. 
Most excitingly, the application to Kitaev materials appears especially promising because the Zeeman anisotropy of the edge Raman signal is a telltale sign of the spin liquid, in particular when combined with Mott-resonant Raman probes.  

\vspace{1cm}
{\bf Acknowledgements:}
We acknowledge support from the Deutsche Forschungsgemeinschaft (DFG, German Research Foundation) under Germany’s Excellence Strategy–EXC– 2111–390814868, DFG grants No. KN1254/1-2, KN1254/2-1, and TRR 360 - 492547816, as well as the Munich Quantum Valley, which is supported by the Bavarian state government with funds from the Hightech Agenda Bayern Plus. JK acknowledges helpful discussions with C.H. Back and C.P. Pfleiderer. AN acknowledges helpful discussions with Rao Peng and Jonas Habel.

{\section{Appendix}}
\emph{Perturbation dependence of the Raman signal---}
We compute the Raman spectra as $h$ varies, keeping $|h/t|$ fixed. We fix a counterclockwise orientation such that $t_{ij} =  -t_{ji} =t>0$ if  $j$ is the closest site to $i$ that also has a positive relative angle to $ i$, otherwise $t_{ij}$ is zero. Then we repeat the calculation, letting $t \rightarrow -t$. The resulting Raman spectra shown in figure \ref{fig:ramanlingrowth} exhibit a strong dependence on the sign of $t$, but also a linear dependence on $h$. 
\begin{figure}
    \centering
\includegraphics[]{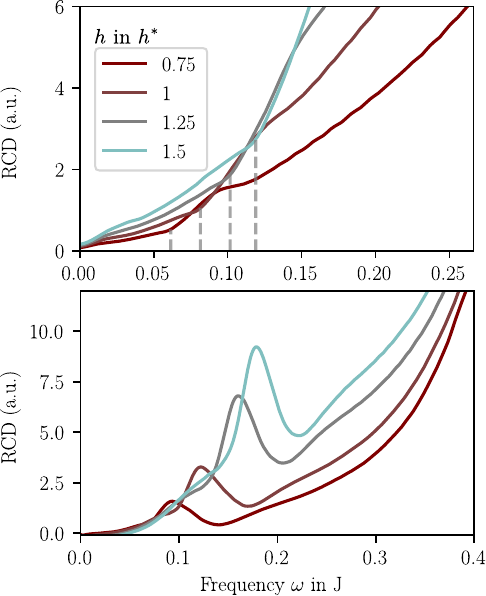}
    \caption{Raman spectra of a KSL for various Zeeman fields $h$ in the presence of a dispersion term $t$ with ratio $|\frac{h}{t}| = 2$ generated by non-Kitaev interactions and $r1=36a$,$r2=46a$, $\kappa=0.12J$, $N = 170$ averaged over $500$ disorder realizations. The upper panel shows the case for positive $t$, the lower panel the case of negative $t$. In the upper panel, we also indicated the frequency where $\partial_\omega I(\omega)$ attains a maximum, demonstrating the linear frequency-dependence of certain features in the RCD to the Zeeman field, in addition to the intensity dependence. In the lower panel, this is better visible from the shift of the local maxima.}
    \label{fig:ramanlingrowth}
\end{figure}\\

\emph{Analysis of matrix elements}---To verify the picture that the Raman signal originates from the mixing of energy scales with different approximate symmetries, we analyze the structure of the Raman transitions.
First, we define $O_{nm} = |\langle n,m| R|\text{GS}\rangle|^2$, which gives a contribution to the Raman signal involving a pair excitation with energy $E_n$ and $E_m$ on top of the ground state $\ket{\text{GS}}$. The total contribution of the state $n$ to the edge Raman signal is
\begin{align}
    K(E_n) = \sum_{m,E_m<\Delta} O_{mn} \label{eq:totaln}.
\end{align}
Then, for a given state of energy $E_n$, we compute the Raman center of energy
\begin{align}
    \text{RCE}(E_n) := \sum_{m,E_m<\Delta} \frac{E_m O_{mn}}{K(E_n)}. \label{eq:RCE}
\end{align}
In figure \ref{fig:ramanlingrowth} we show $\text{RCE}(E_n)$ for the RCD signal computed on a hexagonal flake geometry without gauge-matter mixing and no disorder. Identifying $E_L \approx \Delta/2 \approx 0.2t$, we observe first that low energy modes contribute to the Raman signal with high energy modes and vice versa. Furthermore, $K(E)$ appears to increase with $|E-E_L|$. For energies closer to $E_L$, the RCE is also close to $E_L$, because it receives contributions from both high and low energy modes.

\begin{figure}
    \centering
\includegraphics[]{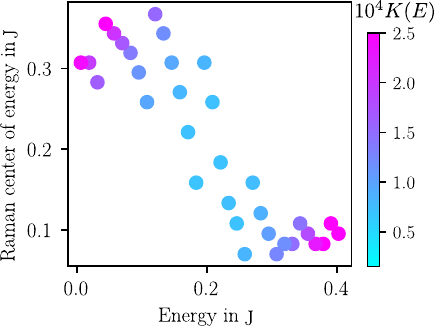}
    \caption{The Raman center of energy \eqref{eq:RCE} for each edge state with energy $E_n$ calculated on the outer boundary of a hexagonal flake with $R=66a, \kappa = 0.08t,h=0$ and no disorder. The color indicates the total contribution to the Raman signal for each mode \eqref{eq:totaln} 
    .}
    \label{fig:ramanlingrowth}
\end{figure}
\emph{Projective angular momentum}---To determine the projective action of lattice rotation $G$ in an arbitrary gauge configuration, we consider its action on the eigenvectors $V$ of the gauge fixed Hamiltonian, with row index in site basis and column index in the eigenbasis of the Hamiltonian.
Since $G$ is realized projectively, we have
$ GV = U_G V D$ where $D$ diagonal in the Hamiltonian eigenbasis and the gauge transformation  $U_G$  is diagonal in the site basis with entries in $\mathbb{Z}_2$.  To find $U_G$ we 
we fix one non-degenerate eigenvector $v$ and determine  $d= \frac{(Gv)_x}{v_x}$ for an arbitrary but fixed site index  $x$, then we compute $(U_G)_y = d^*(Gv)_yv^*_y $. 
 Note that the eigenvalues $D$ transform under the IGG such that eigenvalues related by a $\pi$ phase shift are equivalent. 
 
\begin{figure}
    \centering
    \includegraphics[]{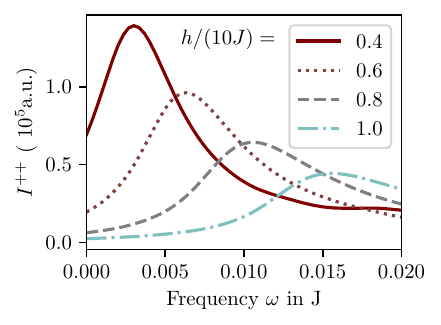}
    \caption{The Raman intensity $I^{++}$ (note the different unit)  calculated with the three spin Raman vertex eq. \eqref{eq:resraman} on the outer edge of disorder free hexagonal disk with $r_2 = 56a$, $r_1= 6a$ for various $\kappa = h$. }
    \label{fig:Resonant Raman}
\end{figure}

 To determine $\Delta l$ for the state $\ket{n,m} = f^\dagger_m f^\dagger_n\ket{GS}$, where $f$ are eigenmodes of the Majorana Hamiltonian, consider that the one-body density matrix of this state
 \begin{align}
     P^{nm}_{ij} &= \bra{n,m} a_i a_j\ket{n,m}
 \end{align}
equals the one-body density matrix of the complex fermion problem with Hamiltonian $\iu A$ and a particle-hole excitation.
Since 
\begin{align}
    \Delta l = \Tr P^{nm} U_G C_6 -\Tr P^{GS} U_G C_6 
\end{align}
The change in angular momentum is given by  the difference of the angular momenta of the $n$ and $m$ eigenvectors of $\iu A$, i.e.,

\begin{align}
    \Delta l =\frac{6}{2 \pi}\log [{\mathcal{Q}^-_m}^\dagger  (U_{C_6} C_6 )^\dagger \mathcal{Q}^-_m  ] [{\mathcal{Q}^+_n}^\dagger  U_{C_6} C_6 \mathcal{Q}^+_n].
\end{align}
\emph{Resonant Raman scattering}---To access the DOS of the bond fermions, we look at the next-leading order Raman operator, which becomes significant once the driving frequency approaches the Mott gap \cite{Brent16, Johannes16}:
\begin{align}
    R_{res} = i\sum_{\langle \langle i (j) k \rangle \rangle,{\alpha \beta \gamma}}M^{\alpha \beta \gamma}_{ijk}  \sigma_i^\alpha \sigma_j^\beta \sigma_k^\gamma \label{eq:resraman}
\end{align}
where $j$ is a common nearest neighbor (NN) to the next NNs $i,k$. In the low-energy sector, the bond variables are chosen as for the three spin term responsible for $\kappa$. However, at the boundary, this operator also induces terms of the form. 
\begin{align*}
    \sum_{i,k \in \mathbf{E}} M^1_{ik} b_i b_k \text{ and } \sum_{i \in \mathbf{E}, k} M^2_{ik} b_i c_k
\end{align*}
where, in the $++$ polarization channel, $M_{ik} = \iu \frac{\sqrt{3}a^2}{6}\lambda_r \nu_{ik}  $ for a microscopically determined $\lambda$ which we set $\lambda_r =\lambda/2$. The resulting Raman spectra, shown fig. \ref{fig:Resonant Raman} show that the signal increases as the Zeeman field is lowered and can exceed the RCD signal by more then five orders of magnitude.  
\\

\begin{figure}
    \centering
    \includegraphics[]{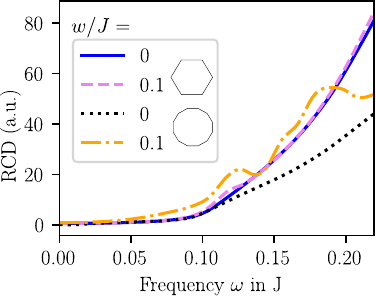}
    \caption{Raman signal calculated with $H^{htw}$ on a hexagonal and dodecagonal geometry with $r_1=16$, $r_2=56$ $h = \kappa = 0.05J$ for $\omega = 0.1J$, and $\omega=0J$.   }
    \label{fig:Gammadep}
\end{figure}

\emph{Zig-zag vs armchair edges}---
We now consider the Hamiltonian
\begin{align}
    H^{htw} = H^{ht}+ w\sum_{\langle  i, j\rangle| i \notin \mathbf{E}, j\in \mathbf{E}} \iu u_{ij}  c_i b_j \label{eq:htw}
\end{align}
where $t_{ij} = 0.1u_{ij} $ for NNN and $t_{ij} = w u_{ij}$ for NN. We note that unlike NNN terms the additional NN terms also contribute to the Raman signal. We consider a hexagonal geometry where the edge is purely of zig-zag type, except at the corners, and a dodecagonal geometry where there is an equal number of zig-zag and armchair type edges. The results are shown in fig. \ref{fig:Gammadep}. We observe that $\omega$ has a negligible influence in the hexagonal geometry while it is significant in the dodecagonal geometry. 

\emph{RCD and chirality}
Next, we discuss the validity of $I^{RCD}$ as a measure of chirality. The chirality, i.e., the direction of propagation and the sign of the RCD are related under the assumption that the edge mode velocity is a non-zero function of angular momentum, the rotation symmetry is continuous and a choice of normal vector that defines the orientation,  such that the sign of $v$,  fixes the sign of $l$ which in turn implies the direction of propagation. E.g., if $v>0$, the edge can only absorb positive $l$, implying a positive sign of the RCD and an anticlockwise direction with respect to the normal vector used to define $l$. There are multiple ways in which these assumptions may be violated. First, consider that the inner edge, which moves clockwise, also gives a positive RCD. Indeed, while the low energy modes on the outer edge transform close to the identity under rotations, the low energy modes at the inner edge transform close to a $\pi$ rotation, and their angular momenta are defined in the tangent space at $-1$ such that clockwise movement goes into the direction of positive $l$. Second, the Chern number protects only the difference between clockwise and anticlockwise movers, so $v(l)$ may cross $0$, and transitions involving states with negative angular-momentum transfer are allowed. Lastly, the sign of the RCD contributions involving modes without an approximate continuous rotation symmetry is not fixed by $v$. Therefore, and as we have also confirmed with our simulations, the sign of the RCD does not allow conclusions on the chirality of the edge mode. In particular, the in-gap RCD can also be finite in the topologically trivial regime of strongly anisotropic $J^\gamma$.
\end{document}